\documentclass[journal=jctcce,manuscript=article,
layout=twocolumn,email=false]{achemso}
\setkeys{acs}{articletitle=true} 
\usepackage{graphicx}
\usepackage{amsmath}
\usepackage{amsfonts}
\usepackage{amssymb}
\usepackage{hyperref}
\usepackage{dcolumn}% Align table columns on decimal point
\usepackage{bm}% bold math
\usepackage{physics}

\SectionNumbersOn

\DeclareUnicodeCharacter{2212}{-}

\usepackage[utf8]{inputenc}
\usepackage[T1]{fontenc}
\usepackage{mathptmx}
\let\oldmaketitle\maketitle
\let\maketitle\relax

\title{Coupled cluster and perturbation theories based on a cluster mean-field reference applied to strongly correlated spin systems}

\author{Athanasios Papastathopoulos-Katsaros}
\affiliation{Department of Chemistry, Rice University, Houston, Texas 77005, USA}

\author{Carlos A. Jiménez-Hoyos}
\affiliation{ Department of Chemistry, Wesleyan University, Middletown, Connecticut 06459, USA}

\author{Thomas M. Henderson}
\affiliation{Department of Chemistry, Rice University, Houston, Texas 77005, USA}
\alsoaffiliation{Department of Physics and Astronomy, Rice University, Houston, Texas 77005, USA}

\author{Gustavo E. Scuseria}
\affiliation{Department of Chemistry, Rice University, Houston, Texas 77005, USA}
\alsoaffiliation{Department of Physics and Astronomy, Rice University, Houston, Texas 77005, USA}

\begin{document}
%-------------------------------------

%% EXTRA COMMANDS FOR ABSTRACT: so that abstract is single column in two-column layout %%

\twocolumn[
\begin{@twocolumnfalse}
\oldmaketitle
\begin{abstract}
We introduce perturbation and coupled-cluster theories based on a cluster mean-field reference for describing the ground state of strongly-correlated spin systems. In cluster mean-field, the ground state wavefunction
is written as a simple tensor product of optimized cluster states. The cluster language and the mean-field nature of the ansatz allows for a straightforward improvement which uses perturbation theory and coupled-cluster to account for inter-cluster correlations. We present benchmark calculations on the 1D chain and 2D square $J_1-J_2$ Heisenberg model, using cluster mean-field, perturbation theory and coupled-cluster. We also present an extrapolation scheme that allows us to compute thermodynamic limit energies accurately. Our results indicate that, with sufficiently large clusters, the correlated methods (cPT2, cPT4 and cCCSD) can provide a relatively accurate description of the Heisenberg model in the regimes considered, which suggests that the methods presented can be used for other strongly-correlated systems. Some ways to improve upon the methods presented in this work are discussed.
\end{abstract}
\end{@twocolumnfalse}
]

%-------------------------------------

\section{Introduction}
\par Many physical systems can be envisioned as weakly correlated collections of objects which individually have large strong (or static) correlations. One obvious example is a two-component chemical system in which the constituents are well separated and weakly correlated with one another, but in which each constituent has significant static correlation (by, for example, having multiple transition metal centers). Another important example consists of systems which have localized spins, such as spin-lattices. In these lattices, the interactions between spins usually decay with respect to the distance and display features of locality. By exploiting the decay of the interaction, one can divide lattices which display this behavior (although not all of them do) into components which for our purposes we call "clusters`` or "tiles``.
\par Unfortunately, despite tremendous effort and good progress, the accurate and efficient description of such systems is an open problem in quantum chemistry, because single-reference methods are generally inadequate.\cite{bulik_can_2015} Approaches based on composite particles (e.g. composite fermion-boson mapping \cite{zhao_composite_2014}) have been proposed to tackle the multi-reference character inherent in these systems. In this work, which is a continuation of the work in Ref.~\citenum{jimenez-hoyos_cluster-based_2015}, we discuss cluster mean-field (cMF) and correlated extensions based on perturbation theory (cPT) and coupled-cluster (cCC) for spin systems. The same formalism can be generalized to fermions, as shown in Ref.~\citenum{jimenez-hoyos_cluster-based_2015} for cMF and cPT2, and Ref.\citenum{fang_block_2007} for cCC. Although the fermionic case will be more complicated, it still remains feasible. These cluster-based methods allow us to tackle the problem at its root, by creating an ansatz that takes into account the cluster-like character of the systems. More specifically, as a first step, we divide the system under study into clusters, which we treat exactly with full configuration interaction (FCI). We account for the inter-cluster interactions in a mean-field manner (cMF). When there are no inter-cluster correlations, cMF is exact. In practice, however, there will generally be correlations between the clusters which must be accounted for. Here, we aim to study methods which can do so. These methods are cluster-based generalizations of traditional quantum chemistry approaches, such as perturbation theory (cPT) and coupled-cluster (cCC). These approaches should be ideally suited to treating systems with strong correlations within each cluster and weak correlations between them.
\par To mathematically write the wavefunction, we use composite many-spin cluster states. Each cluster is defined by a fixed number of neighboring
sites. These clusters can have any shape and any size, although for 2D systems, compact shapes most often provide better results. The cluster states are a subset of all the available many-spin states (most often all the states of the $S_z = 0$ sector). We presume that an accurate zero-th order description of the ground state of the full system can be prepared as a product of cluster states which we variationally determine in the presence of the other clusters. The optimization provides not only the optimal cMF state, but also a renormalized Hamiltonian expressed in term of cluster states. Traditional many-body approaches can then be used, on this renormalized Hamiltonian, to account for the missing inter-cluster correlations.
\par In related work, Isaev, Ortiz, and Dukelsky\cite{isaev_hierarchical_2009} considered a similar ansatz to ours.  Our approach differs from that used in Ref.~\citenum{isaev_hierarchical_2009} in not requiring the individual clusters to share the same ground state, which provides more variational freedom. That is, the ground state of each cluster is optimized independently allowing for (translational and spin) symmetry-broken solutions. In addition, we here consider two common approaches from quantum chemistry (Rayleigh-Schr\"{o}dinger perturbation theory (RS-PT)\cite{schrodinger_quantisierung_1926} and coupled-cluster (CC)\cite{shavitt_many-body_2009}) as a means to obtain a correlated approach defined in terms of clusters. Regarding the cluster-based generalization of traditional perturbation theory, block correlated second order perturbation theory with a generalized valence bond (GVB) reference has been previously proposed by Li and coworkers\cite{xu_block_2013}. The key concepts and the formulation are the same as in this work, although the GVB reference is just a particular cluster mean-field reference. Our coupled-cluster approach is inspired by Li's\cite{li_block-correlated_2004, wang_describing_2020} block-correlated coupled-cluster method. More specifically, the wavefunction ansatz and the definition of cluster operators are the same as those in Ref.~\citenum{li_block-correlated_2004} (a "cluster`` in our work has the same meaning as a "block`` in that work, meaning a subset of orbitals), except that for BCCC the ground state of each cluster was not optimized in the presence of other clusters as we do here. Block-correlated coupled cluster has been used with high success in quantum chemistry to describe strongly-correlated molecular systems using a complete active-space\cite{fang_block_2008, shen_block_2009} reference state. In addition, Abraham and Mayhall\cite{abraham_selected_2020} went beyond cMF by implementing a selective configuration interaction framework and had very accurate results for fermionic systems. Moreover, our approach shares some conceptual similarities with density matrix embedding theory (DMET)\cite{knizia_density_2012, wouters_practical_2016}, active-space decomposition (ASD)\cite{parker_communication_2013}, and localized active space self-consistent field method (LASSCF)\cite{hermes_multiconfigurational_2019}, where the common denominator is the use of clusters (or fragments) and a lower level treatment between those fragments. More specifically, DMET embeds a local fragment in an environment that is treated at a low level, ASD computes a larger fragment's complete-active-space wavefunction while only constructing a smaller fragment's active-space wavefunctions, and in LAS the embedded subsystem comes from dividing the active space into unentangled active subspaces each localized to one fragment. Lastly, a cluster product state is also connected with tensor network (TN) techniques\cite{cirac_renormalization_2009} that have been gaining popularity for treating strongly-correlated systems. For more information on this connection and the advantages of cMF in contrast to other more sophisticated approaches, we refer the reader to Ref.~\citenum{jimenez-hoyos_cluster-based_2015}. 
\par We would like to point out a few useful aspects of cMF. First of all, straightforward symmetry breaking ($S^2$ symmetry) can partially account for inter-cluster correlations, much as in the mean-field description of fermions. In the system we will be considering, for example, inter-cluster correlations can be approximately described by breaking the $S^2$ spin symmetry. Moreover, cMF is exact in the limit that we have only one cluster, while for systems with $n$ clusters, cCC with $n$-fold excitations is exact so long as we consider every state in every cluster. When each cluster contains one particle, cMF is Hartree-Fock (HF), cPT is conventional perturbation theory and so on. This allows us for a straightforward comparison to conventional single-reference methods.

\par Our objective with this work is two-fold. First, we
present the cMF formalism and provide details of
the RS-PT (cPTn) and the coupled-cluster formulation we use (cCCSD). Our second objective is to apply these techniques to a simple strongly-correlated system: the Heisenberg model in a 1D chain and 2D square lattice and verify whether clusterization is useful.

\par Spin lattices and more specifically, Heisenberg models, are of significant chemical importance. For example, iron-sulfur clusters relevant to nitrogen fixation or photosynthesis, such as ferredoxins, have been treated according to the Heisenberg model.\cite{chan} Single molecule magnets have possible applications in quantum computers as the smallest practical unit for magnetic memory. These molecules are usually metal clusters, and the magnetic coupling between the spins of the metal ions can also be described by a Heisenberg Hamiltonian.\cite{molecularmagnets} Lastly, electrides, conjugated hydrocarbons and a few superconductors have some of their features modelled after Heisenberg exchange interactions.\cite{apps1,apps2,apps3} The Heisenberg model serves as a useful benchmark system since it has a variety of phases and has been well studied, but has many fewer degrees of freedom than a fermionic system with the same number of sites.
\par We should emphasize that the focus of our work is on developing a method for strongly-correlated systems. While we have chosen to benchmark our techniques on the Heisenberg model for the reasons we have already explained, it is not our intention here to significantly advance our understanding of Heisenberg systems. It is important to underline that the 1D case of the Heisenberg model is exactly solvable,\cite{bethe_zur_1931} and Li's block-correlated coupled-cluster method is very accurate for the 1D and the 2D ladder case, so even though our purpose is to show whether clusterization is useful for any general Hamiltonian, a big part of our focus is on the general 2D model. This model has received numerous studies in the past two decades, using various
methods such as exact diagonalization,\cite{dagotto_phase_1989,schulz_finite-size_1992, richter_spin-12_2010, capriotti_spontaneous_2000, mambrini_plaquette_2006, schulz_magnetic_1996} coupled-cluster,\cite{schmalfus_quantum_2006, darradi_ground_2008, bishop_phase_1998, richter_spin-12_2015, bishop_main} density-matrix renormalization group (DMRG),\cite{jiang_spin_2012, gong_plaquette_2014} matrix-product or tensor-network based algorithms,\cite{murg_exploring_2009, yu_spin-_2012, wang_constructing_2013} resonating valence bond (RVB)\cite{capriotti_resonating_2001} and quantum Monte Carlo (QMC).\cite{sandvik_finite-size_1997} Even though the computational cost of cMF and exact diagonalization are of the same order of magnitute, in contrast to exact diagonalization, cMF allows us to find different solutions for different regimes and we can also use less costly approximate methods in each cluster instead to reduce the scaling of cMF. We compare our results with calculations from Refs.~\citenum{richter_spin-12_2015}, which we use as a reference. Those results are obtained using high-order single-reference coupled-cluster, and are regarded among the most accurate results for the specific system (more information is available in the papers mentioned) and therefore we consider them useful for a quantitative comparison. Our results show that cPT2, cPT4 and cCCSD significantly improve upon cMF and can provide an accurate description of the ground state of the square $J_1$-$J_2$ Heisenberg lattice. 
\par The rest of the article is organized as follows. In section \ref{2.0} we present the formalism behind cMF, cPT and cCCSD. Section \ref{3.0} provides some practical computational details regarding the calculations presented in this work.
In section \ref{4.0} we present the results of cMF, cPT2, cPT4 and cCCSD calculations for the 2D square $J_1-J_2$ Heisenberg model. A brief discussion following the results is presented in section \ref{5.0}, as well as some ideas as to how to improve the approaches presented here. Lastly, section \ref{6.0} is dedicated to some general conclusions.

\begin{figure}
\centering
\includegraphics[scale=0.4]{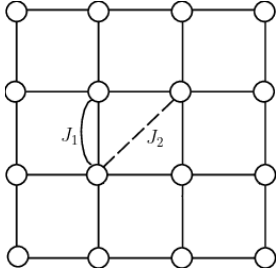}
\caption{Nearest-neighbor ($J_1$) and next-nearest neighbor ($J_2$) interactions.}
\label{j1j2}
\end{figure}

\section{Formalism}\label{2.0}
\subsection{Heisenberg model}
In this work, we focus our attention on the $J_1$ - $J_2$ Heisenberg model in two-dimensions on a square grid. The Heisenberg model describes a collection of spins in a lattice (of finite size $N$) interacting through the Hamiltonian 
\begin{eqnarray}
H = J_1 \sum_{\langle ij \rangle} \vec {S_i} \cdot \vec {S_j} + J_2 \sum_{\langle \langle ij \rangle \rangle} \vec {S_i} \cdot \vec {S_j}
\end{eqnarray}
where $\vec {S_i}$ is the spin-$\frac{1}{2}$ operator on site $i$, $J_1$ and $J_2$ are the nearest-neighbor and the  second-nearest neighbor coupling coefficients respectively (see Fig.~\ref{j1j2}), and the notation $\langle ij \rangle$ implies interaction among nearest-neighbors, while $\langle \langle ij \rangle \rangle$ implies interaction among next-nearest neighbors. In the following, we confine ourselves to the antiferromagnetic (AFM) case $J_1, J_2 > 0$.
\par
\begin{figure}
\centering
\includegraphics[scale=0.7]{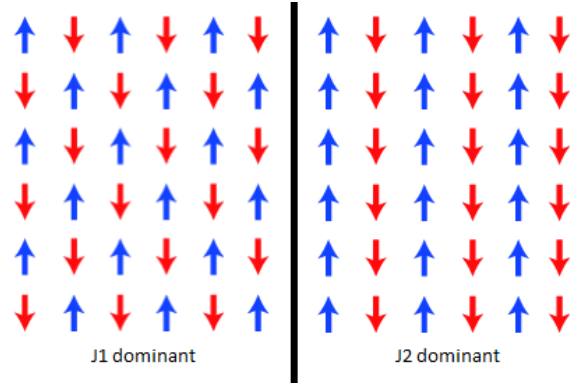}
\caption{Néel (left) and collinear (right) antiferromagnetic phases of the square $J_1 - J_2$ Heisenberg model. In between, there is a non-magnetic phase whose precise form is debated.}
\label{phases}
\end{figure}

 As mentioned previously, this model has been studied extensively in the past two decades, using various
methods. It has been established that in the regime $0 \lesssim J_2/J_1 \lesssim 0.4$, the ground state is an AFM phase with Néel order, due to the dominance of the nearest-neighbor interactions $J_1$. In $J_2/J_1 \gtrsim 0.6$, the ground state displays an AFM phase with collinear long-range order character due to the dominance of the next-nearest-neighbor coupling $J_2$ (see Fig.~\ref{phases}). In the regime $0.4 \lesssim J_2/J_1 \lesssim 0.6$, which we refer to as the "paramagnetic phase``, the system is frustrated and the Néel and the collinear orders compete. The nature of this intermediate ground state is still a much debated issue,\cite{schulz_finite-size_1992, gelfand_series_1990,zhitomirsky_valence-bond_1996,takano_nonlinear_2003, isaev_hierarchical_2009, lante_ising_2006, jiang_spin_2012, wang_constructing_2013, hu_direct_2013, li_gapped_2012} as are the type of the phase transitions and the transition points. In this work, we focus mostly on the convergence of the energy to the exact result in each of the separate regimes, and we only pay minor attention to the correct location of the critical points. We need to mention, though, that there is a second-order transition from the Néel to the paramagnetic phase, whereas there is a first-order transition from the paramagnetic to the collinear antiferromagnetic phase.

\subsection{Cluster Mean-Field}
Our formalism of cluster mean-field (cMF) is based on Ref.~\citenum{jimenez-hoyos_cluster-based_2015}, but in this paper we confine ourselves to the spin case, which consists of a subset of configurations found in fermionic systems. For more details we refer the reader to that work, but below we present the general framework.

\par Let the lattice states be grouped, according to some criterion (such as proximity in real space), into clusters of size $l_1, l_2, . . . l_n$, where $n$ is the number of such clusters. Formally, the Hilbert space of each cluster is of size $2^{l_i}$, as in each site we can either have a spin-up or a spin-down. We choose to work with eigenstates of $S_z$ in each cluster, thereby reducing the effective dimension of the Hilbert space. Similarly to Ref.~\citenum{jiang_spin_2012}, the Hilbert space of the full system is simply given by the tensor product of the Hilbert spaces of all clusters. 

\par A second-quantized formulation in terms of cluster product states can also be established. Let $A^{\dag}_{I,c} (A_{I,c})$ create (annihilate) the $I$-th state in cluster $c$. This $I$-th state is a linear combination of many-spin basis states (possibly mixing states with different $S_z$) constructed as products of the states in the cluster. We formally write
\begin{eqnarray}
\ket{I}_c = A^{\dag}_{I,c}\ket{-}_c
\end{eqnarray}where $\ket{-}_c$ is the vacuum state in cluster c, a useful abstract construct.

\par Each cluster product state is formally built as
\begin{eqnarray}
\ket{I}_1\ket{J}_2 . . .\ket{Z}_n \equiv A^{\dag}_{I,1}A^{\dag}_{J,2} . . . A^{\dag}_{Z,n} \ket{-}
\end{eqnarray}
where $\ket{-}$ is a vacuum state for the full system.
In this work as a starting point, we consider a cluster product (mean-field) state as a variational ansatz for the ground state reference wavefunction, on which we will later build correlation. That is, the ansatz $\ket{\Phi_0}$ for the reference state is given by
\begin{eqnarray}
\ket{\Phi_0} = \ket{0}_1\ket{0}_2 . . . \ket{0}_n
\end{eqnarray}
where the 0 label indicates the ground state of each cluster in the presence of the other clusters. The lowest energy cMF state is obtained by a variational minimization scheme, as outlined in Ref.~\citenum{jimenez-hoyos_cluster-based_2015}. 

\par Defining excited configurations is straightforward. We can write them as
\begin{align}
\ket{\Phi_{Ii}} &= \ket{0}_1...\ket{I}_i...\ket{0}_n, \\
\ket{\Phi_{Ii;Jj}} &= \ket{0}_1...\ket{I}_i...\ket{J}_j...\ket{0}_n, \\
\ket{\Phi_{Ii;Jj;Kk}} &= \ket{0}_1...\ket{I}_i...\ket{J}_j...\ket{K}_k...\ket{0}_n 
\end{align}
for singly-, doubly-, and triply-excited clusters.

\par Before proceeding further, let us comment on the nature of the cluster product states considered in this work. We indicated above that the ground state of each cluster is expressed as a linear combination of the many-spin basis states in it. The expansion over those states can be restricted and for our purposes, we confine ourselves to cases with a specific $S_z$. This is done in order for the cluster product state $\ket{\Phi_0}$ to be an eigenfunction of $S_z$ and to reduce the dimension of the ground state vector in each cluster. Lastly, we choose to break $S^2$ in each cluster to develop long-range magnetic ordering.

\subsection{Matrix elements and cMF optimization}
The evaluation of the matrix elements is again similar to Ref.~\citenum{jimenez-hoyos_cluster-based_2015}, but an important difference that we need to point out is that the spin cluster Hamiltonian has 1- and 2-cluster elements but not 3- or 4-cluster elements. This difference arises because of the form of the Hamiltonian, which does not have more than 2-spin interactions. This significantly simplifies the procedure for both the cMF optimization, as well as the cPT and cCC extensions mentioned later. For details regarding the cMF optimization, we refer the reader to Ref.~\citenum{jimenez-hoyos_cluster-based_2015}. We note that the zero-th order cluster Hamiltonian for a given number of spins is given by
\begin{eqnarray}
 H_c^0 = \sum\limits_{ij\in c}J_{ij} ({\vec {S_{ij}} }\cdot {\vec{ S_{ij}}}) + \sum_{i \in c} \sum_{j \not \in c} J_{ij} ({\vec{ S_{ij}}} \cdot \langle {\vec{ S_{ij} }}\rangle)
\end{eqnarray}
where $J_{ij}$ refers to the $i$ and $j$ neighbor coupling coefficient. We choose to perform the optimization  self-consistently, in order to minimize the energy. 
\par We think one final comment contrasting cMF and standard diagonalization techniques is necessary. Let us focus on the case of a finite lattice with periodic boundary conditions. The scaling of cMF with respect to cluster size is similar to that of exact diagonalization performed on a full lattice of the same size as the cluster, though with a larger prefactor since the equations are solved self-consistently and the ansatz explicitly breaks the translational (and, possibly, spin) symmetry of the lattice. Thus, we can reach cluster sizes comparable to those achievable with exact diagonalization results (though of course cMF can have many clusters). In this work, the largest cMF calculation reported used a $6 \times 6$ cluster, and the length of the eigenvector in each tile is $\sim 10^{10}$.

\subsection{Perturbation theory}
Once again, the theory is very similar to Ref.~\citenum{jimenez-hoyos_cluster-based_2015}, but significantly simplified. Generally, in RS-PT, the second-order correction to the ground state energy is evaluated as
\begin{eqnarray}
E^{(2)} = \sum\limits_{\mu \neq 0} \frac{|\hat V_{0\mu}|^2}{\epsilon_0 - \epsilon_{\mu}}
\end{eqnarray}
where $\hat V = \hat H - \hat H_0$ and $V_{0\mu} = \bra{\Phi_0} \hat V \ket{\mu}$. Here, $\mu$ labels the eigenstates of $\hat {H_0}$ and $\epsilon_{\mu}$ are the corresponding
eigenvalues. The excited states framework was explained in section B. As mentioned earlier, the evaluation of the matrix elements is easier compared to the fermionic case, because there are no 3- and 4-cluster interactions, so the cost of cPT2 is $\mathcal{O}(n^2K^2)$, where $K$ is the number of excited states in each cluster. We would like also to remind the reader that we use all the states in the Hilbert space of each cluster (although in practice for cPT2 we only need states with $S_z = m, m+1, m-1$, where $m$ is the $S_z$ value used in cMF), even though in the cMF optimization only one $S_z$ sector of the Hilbert space was considered. In addition, due to the structure of the Hamiltonian, we only have two types of 2-tile  interactions ($S_z$-preserving and $S_z$ mixing), as opposed to 5 (see Ref.~\citenum{jimenez-hoyos_cluster-based_2015}) in the fermionic case. Lastly, in order to go beyond second order, we employed the usual scheme of RS-PT.

\subsection{Coupled-cluster theory}
In this paper, our aim is to go beyond perturbation theory to a coupled-cluster framework. Our work bears connection to Ref.~\citenum{li_block-correlated_2004}, but it is important to note that in that work, the ground state of each cluster was  neither optimized in the presence of other clusters nor tested on the same model.
\par The cluster coupled-cluster (cCC) expansion of the ground-state wave function can be written in an intermediate normalized form as follows:
\begin{eqnarray}
\ket{\Psi} = e^{\hat T}\ket{\Phi_0}
\end{eqnarray}
where $\hat T$ is the cluster operator. In this work, we focus our attention to singles and doubles, therefore 
\begin{eqnarray}
\hat T = \hat T_1 + \hat T_2
\end{eqnarray}
and in the cluster language, $\hat T_1$ and $\hat T_2$ operators can be written as
\begin{align}
\hat T_1 &= \sum_i \sum_{I(i)} t_{I_i} A^\dag_{I_i} A_{0_i} \\
\hat T_2 &= \frac{1}{2!} \sum_i \sum_{j \neq i} \sum_{I(i)} \sum_{J(j)} t_{I_i J_j} A^\dag_{I_i} A_{0_i} A^\dag_{J_j} A_{0_j}
\end{align}
where the coefficients $t_{I_i}$ and $t_{I_i J_j}$ are the single and double amplitudes respectively, the operator $A^\dag_{I_i}$ excites cluster $i$ to state $I$ and operator $A_{0_i}$ de-excites cluster $i$ to its ground state. Also, we deem necessary to emphasize the connection between $A$ and the excitation operators of single-reference coupled-cluster; both of them share the same properties, meaning that they are nilpotent and have the same commutation relations.
\par By projecting onto $\ket{\Phi_0}$ and the space of singly and doubly excited configuration functions, and by utilizing the fact that for spin systems we have up to 2-cluster excitations, the cCCSD equations naturally truncate and become \\

\begin{subequations}
\begin{align}
E_{cCCSD} =  \bra{\Phi_0}&H\ket{\Phi_0} +  \bra{\Phi_0}H\ket{T_1\Phi_0} \\ & + \nonumber \bra{\Phi_0}H\ket{(T_2+\frac{1}{2}T_1^2)\Phi_0} \\ \nonumber
\end{align}
\begin{align}
E_{cCCSD} \: t_{I_i} = & \bra{\Phi_{I_i}}H\ket{\Phi_0} + \bra{\Phi_{I_i}}H\ket{T_1\Phi_0} \\ + & \nonumber \bra{\Phi_{I_i}}H\ket{(T_2+\frac{1}{2}T_1^2)\Phi_0} \\ + & \nonumber \bra{\Phi_{I_i}}H\ket{(T_2T_1+\frac{1}{6}T_1^3)\Phi_0} \\ \nonumber
\end{align}
\begin{align}
E_{cCCSD} (t_{I_iJ_j}+t_{I_i}t_{J_j}) = \\ \nonumber \bra{\Phi_{I_i;J_j}}&H\ket{\Phi_0} \\ +  \nonumber \bra{\Phi_{I_i;J_j}}&H\ket{T_1\Phi_0} \\ +  \nonumber \bra{\Phi_{I_i;J_j}}&H\ket{(T_2+\frac{1}{2}T_1^2)\Phi_0} \\ + \nonumber \bra{\Phi_{I_i;J_j}}&H\ket{(T_2T_1+\frac{1}{6}T_1^3)\Phi_0} \\ + \nonumber \bra{\Phi_{I_i;J_j}}&H\ket{(\frac{1}{2}T_2^2 + \frac{1}{2}T_2T_1^2+\frac{1}{24}T_1^4)\Phi_0} \nonumber
\end{align}
\end{subequations}
where eq. 14a is the equation for the energy, and 14b and 14c are for the single and double amplitudes, respectively. If we insert eq. 12 and eq. 13, we get a set of nonlinear equations and like in traditional coupled-cluster, we solve the equations self-consistently.
\par It should be emphasized that in a cCCSD calculation, the
computation of the last term in the right-hand side of eq.14c is the most time consuming step, which makes the cCCSD method computationally a $\mathcal{O}(n^4)$ procedure, where $n$ is the number of clusters.
\par Lastly, similar to the conventional truncated CC expansion, the truncated cCC expansion is also size extensive. For more information on the proof, we encourage the reader to check Ref.~\citenum{li_block-correlated_2004}. 
\section{Computational details}\label{3.0}
The cMF, cPT and cCCSD calculations presented in this work
were carried out with a locally prepared code. In all the calculations with even number of spins in each cluster, we use the same number of up and down spins so that the clusters are $S_z$ eigenfunctions with $S_z = 0$, whereas in all the cases with odd number of spins, we use the suitable number of up and down spins in order to be able to construct Néel and collinear antiferromagnetic phases respectively. In this work, we choose to work with eigenstates of $S_z$ in each cluster for computational convenience. This is not a symmetry, so relaxing this constraint could lead to lower variational cMF solutions, but in the Néel and collinear phases it is adequate (the broken $S_z$ states after the optimization will turn back to $S_z$ eigenstates). The full relevant $S_z$ sector of Hilbert space within each cluster was used in constructing the cluster ground state
$\ket{0}_c$. For small cluster sizes, the ground state in each cluster was found by a standard diagonalization of the local
cluster Hamiltonian. For larger cluster sizes, a Davidson\cite{davidson_iterative_1975} algorithm was used to solve for the ground state. In the cPT and cCCSD, we included all the states of all $S_z$ sectors (although in principle the number of states can be truncated, which was done in Ref.~\citenum{jimenez-hoyos_cluster-based_2015}). This means that for a system that is composed of 2 clusters, cCCSD would be exact. For solving the cCCSD equations, the traditional CC iteration scheme was used.\cite{scuseria_accelerating_1986}

\section{Results}\label{4.0}
In this section we present results of cMF, cPT2, cPT4 and cCCSD calculations on the 2D Heisenberg model. We start by
providing the basic idea in section \ref{4.1}, where we
dive into some details regarding the optimization
of cMF states, as well as the calculations of cPT and cCCSD. In addition, we outline the notion of the thermodynamic limit in our cluster-based framework. In section \ref{4.2} we show and compare the results from cMF, cPT2 and cCCSD computations at $J_2/J_1=0$ to accurate numerical estimates of Refs.~\citenum{richter_spin-12_2015} and \citenum{sandvik_finite-size_1997}. In section \ref{4.3}, we show and compare our results for all the phase diagram to results of Ref.~\citenum{richter_spin-12_2015}. Some numerical results are provided for reference in the Supporting Information.
\subsection{Basic idea}\label{4.1}
In this section we discuss most of the aspects regarding cMF, cPT and cCCSD. In this way, we hope that the results presented in subsequent sections will become more transparent to the reader. We consider rectangular Heisenberg periodic lattices with $J_1, J_2 \geq 0$. As a first step, we choose the corresponding tiling scheme (i.e., defining how the spins are grouped into clusters). The optimized cMF state in the cluster is expressed as a linear combination of all the  possible (combinatorial number) resulting configurations and it is optimized by a self-consistent diagonalization of the appropriate cluster Hamiltonian. 
\par At this point, we need to mention that we are interested in the thermodynamic limit (TDL) properties (very large system sizes). The two relevant parameters with which this can be achieved are the cluster size $l$ and the number of clusters $n$; the TDL is when $l \times n \rightarrow \infty$. One way to reach it is to use a fixed cluster size $l$, while taking $n \rightarrow \infty$. In this case, neither cMF nor cCC with a fixed excitation level is exact in the TDL (but cCC using $n$-fold excitations is). A second way to reach the TDL is to directly take cluster size $l \rightarrow \infty$. In this case, cMF is already exact in the TDL, and of course cPT and cCC are likewise. For this work, we used a limited number of cluster sizes, because the exact diagonalization in large clusters becomes very expensive (recall that we require not only the ground state in each cluster but also, in principle, all excited states) and have extrapolated cMF and cPT2 to the TDL by taking $l \rightarrow \infty$; these methods should both give the exact TDL result. We were unable to do so for cCC since we could not perform cCC calculations with sufficiently large tiles to do this extrapolation.  We have instead taken cCC calculations with fixed $l$ and taken the number of tiles to infinity for cCCSD. This gives a TDL estimate, but will not give the exact TDL result.  However, to the extent that the effects of 3-cluster and higher terms are small, the thermodynamic limit of cCCSD should be a reasonable albeit inexact approximation to the exact TDL result (see section \ref{4.2} and section \ref{4.3}).
\par
\begin{figure}
\centering
\includegraphics[scale=0.7]{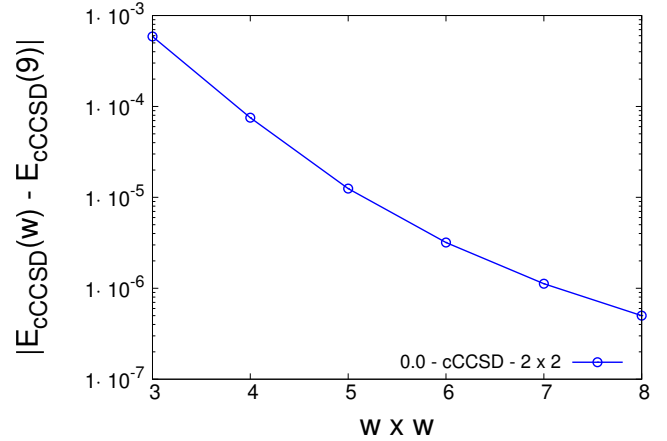}
\caption{cCCSD energies per site using different square cluster configurations. $w$ corresponds to the number of clusters in each dimension of the system. $E_{cCCSD}(9)$ corresponds to cCCSD using $2 \times 2$ tiles in a $18 \times 18$ system ($9 \times 9$ in terms of tiles). 0.0 - cCCSD - $2 \times 2$ denotes cCCSD using $2 \times 2$ tiles with $J_2/J_1=0$. Similar results can be obtained for different values of $J_2/J_1$.}
\label{tdl}
\end{figure}

 Lastly, it is important to note that we can define a ``thermodynamic limit"  system size $L$ corresponding to a specific cluster size $l$ as the the smallest system size for which the results do not change if we increase it further. It is rigorously shown that for the Heisenberg Hamiltonian with periodic boundary conditions and clusters of even dimensions, the thermodynamic limit system size of cMF is twice as large as the cluster size in each dimension of the rectangle, because the cMF solution is uniform (although we do not put it as a constraint, the ground state of cMF ends up as the same wavefunction on every cluster) for the specific model, and for cPT2 it is three times larger, because cPT2 only correlates neighbor tiles; there are no Hamiltonian matrix elements between separated tiles. For cCCSD, though, the above observations do not apply and it is more complicated to compute this limit. Because of this, we consider the cCCSD thermodynamic limit system size as the system size for which the total energy changes after 5 decimal places if we increase the system size any further. In our case, for $2 \times 2$ clusters it was shown that (see Fig.~\ref{tdl}) this can be achieved in a $5 \times 5$ system in terms of clusters, which is equivalent to a 10x10 system in terms of sites. 

\subsection{cMF, cPT2 and cCCSD results at {$J_2/J_1=0$}}\label{4.2}
 All calculations in this section were performed in periodic rectangular lattices. Only uniform tiling schemes were considered; clusters were rectangles of $l$ lattice sites, each filled with $l$ spins. All the results are taken at the thermodynamic limit $L \rightarrow \infty$ for a fixed $l$, as explained in the previous section. We note that broken-symmetry cMF solutions can be achieved, that is, a non-zero magnetization develops on each lattice site (for readers not familiar with spin systems, magnetizations are equivalent to spin densities). Specifically for $J_2/J_1=0$, the ground state of the infinite lattice is a Néel antiferromagnet (see Introduction). Regarding cPT2 and cCCSD, all the computational parameters are the same as for the cMF results. We remind the reader that all the possible excited states were used to compute both the cPT2 and cCCSD energy, except for the 16-site cases (only applicable to cPT2, because we only used $2 \times 2$ clusters with cCCSD) where fewer states \footnote{1250 states in each relevant Sz sector were used. More specifically, 1250/12870 states for $S_z = 0$, 1250/11440 for $S_z = +1$, 1250/11440 for $S_z = -1$  and 0 states in other sectors, as the matrix elements vanish in that case. As a result, the total number of states used is 3750 out of 65536 states in the Hilbert space.} were used; in this case, we ensured that the energy was converged to at least 4 decimal places. The criterion for choosing those states was their corresponding eigenvalues, from which we chose the lowest ones.
 \par In this work, we also went one step beyond and tried to extrapolate the cMF energy to the thermodynamic limit ($L \rightarrow \infty$ and $l \rightarrow \infty$). To do so, we extrapolated $X \times Y$ clusters to the limit $X \rightarrow \infty$, $Y \rightarrow \infty$ by first extrapolating $Y \rightarrow \infty$ for fixed $X$, then extrapolating $X \rightarrow \infty$ from there. 
 
\begin{figure}
\centering
\includegraphics[scale=0.7]{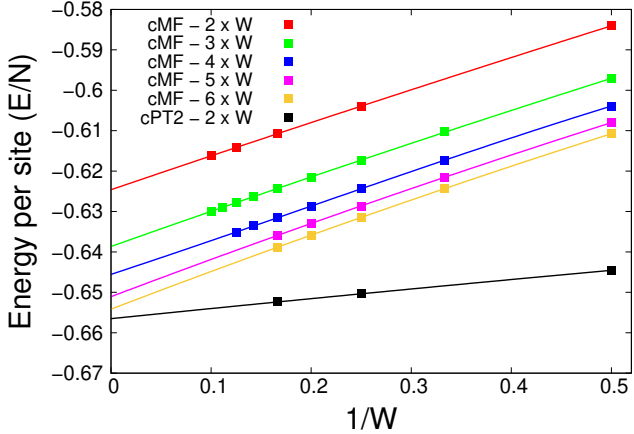}
\caption{Energy per site obtained in cMF and cPT2 calculations at $J_2/J_1 = 0$ using $X \times W$ tiles (with $X$=2,3,4,5,6) as a function of $1/W$.}
\label{2n}
\end{figure}

\begin{figure}
\centering
\includegraphics[scale=0.7]{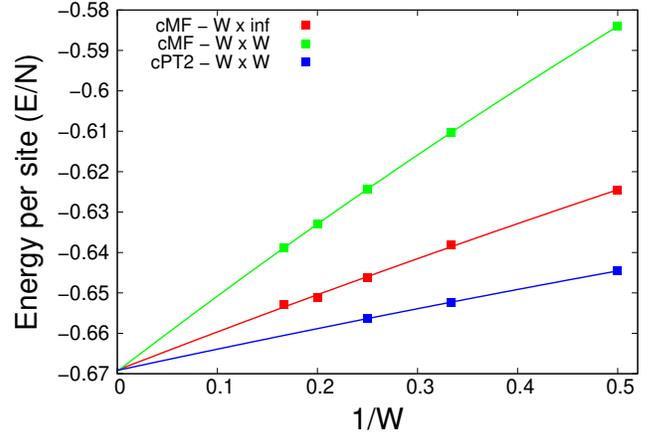}
\caption{Energy per site obtained in cMF and cPT2 calculations for $J_2/J_1 = 0$ as a function of $1/W$. $W \times \infty$ energies refer to the extrapolated energies depicted in Fig.~\ref{2n} and $W \times W$ refer to energies computed with square clusters. The extrapolated results for cMF at ($1/W$ = 0) were obtained by fitting a quadratic polynomial and can be compared with reference calculations from Ref.~\citenum{richter_spin-12_2015} (see Tab.~\ref{tab.1}.) }
\label{nn}
\end{figure}

 Although the energies are asymptotically linear in $1/W$, \footnote{The difference between a $2 \times 6$, a $2 \times 8$ and a $2 \times 10$ tile lies in the addition of extra ''internal`` sites in the tile, i.e., the boundaries have very similar magnetizations. If this is the case, then the error in the energy of the $2 \times W$ vs $2 \times \infty$ should behave nearly linearly, as observed in the plot. On the other hand, the nearly linear nature of the $W \times W$ extrapolation with respect to 1/$W$ can be explained because the error is proportional to the surface (or in this case the perimeter) of the cluster.} we use a quadratic extrapolation since the cluster sizes are not sufficiently large to display the strictly linear asymptotic behavior expected. Figures \ref{2n} and \ref{nn} indeed show a nearly linear behavior as the size of the cluster increases, with some small deviation for larger $1/W$. Lastly, we want to emphasize that as it can be seen from Figs.~\ref{2n} and \ref{nn}, cPT2 converges more rapidly to the TDL than does cMF.
 \par The $J_2/J_1$ = 0 limit has been well studied with careful calculations carried out by several methods and we chose to compare our results to Refs.~\citenum{richter_spin-12_2015},\citenum{darradi_ground_2008},\citenum{sandvik_finite-size_1997}. Our results, as well as the results of the aforementioned methods, are summarized in Tab.~\ref{tab.1}. It is shown that they generally agree to 0.001 x $J_1$. 
 
\begin{table}[h!]
\centering
 \begin{tabular}{|c | c|} 
 \hline
 Clusters & cMF (E/N)\\ [0.5ex] 
 \hline
 $2 \times \infty$ & -0.62445 \\ 
 \hline
  $3 \times \infty$ & -0.63836 \\ 
 \hline
  $4 \times \infty$ & -0.64584 \\ 
 \hline
 $5 \times \infty$ & -0.65035\\
 \hline
  $6 \times \infty$ & -0.65341 \\ 
 \hline
  cMF - $\infty \times \infty$ (1) & -0.66911\\ 
 \hline
  cMF - $\infty \times \infty$ (2) & -0.66940\\
 \hline
  cPT2 - $\infty \times \infty$ (2) & -0.66917\\
 \hline
  CCM Ref.~\citenum{richter_spin-12_2015} & -0.66923\\
  \hline
  QMC Ref.~\citenum{sandvik_finite-size_1997} & -0.66944\\[1ex] 
  \hline
\end{tabular}
\caption{Energy per site obtained by extrapolating cMF calculations at $J_2/J_1=0$. $\infty \times \infty$ (1) is the result from  extrapolating \\ $W \times \infty$ $\rightarrow$ $\infty \times \infty$, and $\infty \times \infty$ (2) is the result from extrapolating \\ $W \times W$ $\rightarrow$ $\infty \times \infty$. The results from coupled-cluster (CCM) and quantum Monte Carlo (QMC) methods are shown for comparison. The agreement is accurate to about 0.001 x $J_1$. }
\label{tab.1}
\end{table}

\begin{figure}
\centering
\includegraphics[scale=1.2]{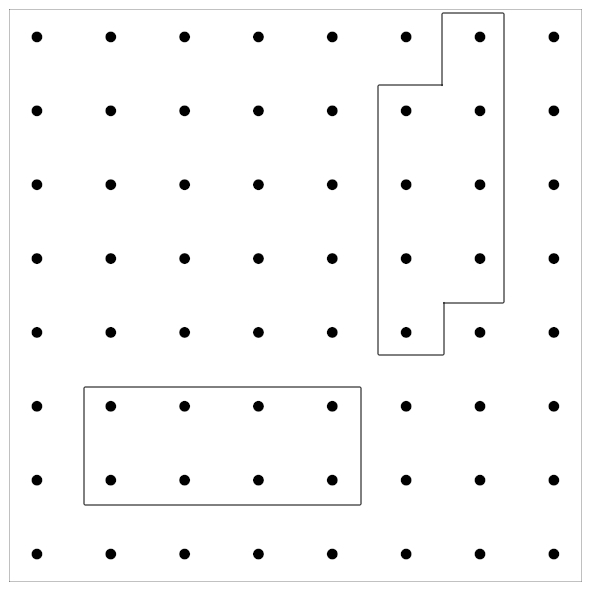}
\caption{$4 \times 2$ (and $2 \times 4$) rectangular (on the left) and tilted (on the right) clusters are shown. Tilted tiles were only used in the collinear antiferromagnetic phase.}
\label{tiles}
\end{figure}

\subsection{cMF, cPT2, cPT4 and cCCSD results for all the phase diagram}\label{4.3}
\begin{figure}
\centering
\includegraphics[scale=0.7]{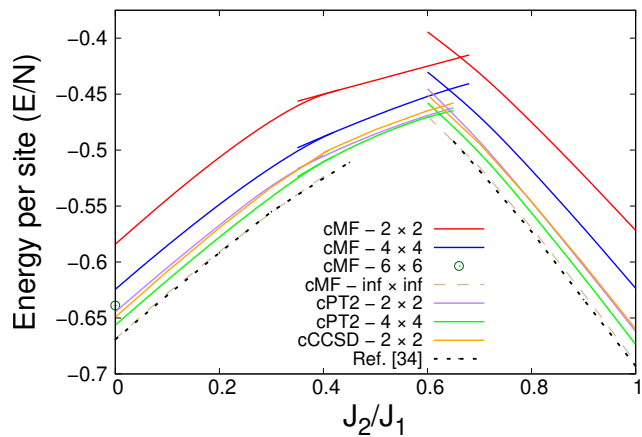}
\caption{Energies per site using different tilings and level of treatment, as well as extrapolated energies computed as described in section \ref{4.2}.}
\label{one_figure}
\end{figure}

We continue by showing our results for $0 \leq J_2/J_1 \leq 1.0$. It is important to remember that there is a Néel phase for $J_2/J_1 < 0.4$, a collinear antiferromagnetic phase for $J_2/J_1 > 0.6$, and a non-magnetic phase for $0.4 < J_2/J_1 < 0.6$. For the collinear phase, tilted lattices as shown in Fig.~\ref{tiles} were used, because they had a more robust behavior when performing the extrapolations. Again, as previously, the same extrapolations schemes were used. In Fig.~\ref{one_figure} we demonstrate the energy per site obtained for different values of $l$, $J_2/J_1$ and the best cMF extrapolation, using cMF, cPT2 and cCCSD. We have also included, for comparison, the extrapolated results to the thermodynamic limit calculated from coupled-cluster calculations computed in Ref.~\citenum{richter_spin-12_2015}. It is important to note that in the intermediate regime we do not show any results, because to our knowledge, there is no reliable extrapolation scheme. Some of the cMF results presented have also been published in Ref.~\citenum{ren_cluster_2014}, but they correspond to fewer cluster sizes and there were no extrapolations performed. We show 3 curves because there are 3 different mean field solutions. Crossing points can be characterized as critical points.
\par Regarding the cMF results, we emphasize the significance of the choice of the cluster, as the $2 \times 2$ cluster cMF results are significantly different from all the other results, which suggests that at the cMF level, $2 \times 2$ clusters are not sufficiently large to capture all the physics of the system. This evidently is not the case for the correlated methods (cPT2 and cPT4). Unfortunately, we currently only have a prototype cCCSD
code which becomes too expensive for clusters larger than $2 \times 2$, but we believe that because the cPT2 and cPT4 results show that $2 \times 2$ is enough, this should similarly be the case for cCCSD. Also, fig.~\ref{one_figure} depicts a subtle difference between the $2 \times 2$ and the $4 \times 4$ cases; it seems that the first-order critical point which is at $\sim$0.6, shifts significantly towards the left as the cluster size increases, while the second-order transition point, which is at $\sim$0.4, stays relatively constant. Lastly, one important observation to point out just for the paramagnetic phase is that the solution is exactly equivalent to the energy of a single tile with open boundary conditions: i. e., the inter-cluster interaction vanishes as there is no  magnetization along the cluster boundaries.
\par Regarding cPT2, we can see that even though it is not variational, there is evidence that the exact energy is lower than the cPT2 energy, because of the results of Refs.~\citenum{richter_spin-12_2015}. There are three important observations. First, increasing the cluster size is not as important for cPT2 as for cMF (see also Fig. \ref{nn}). This can be useful for real systems, because we do not have to use large clusters, whose cost can be prohibitive. We have to underline, however, that cPT2 and cCCSD should converge to the exact answer as the size of the cluster increases. Second, the energy improves significantly even for the $2 \times 2$ case, which suggests that a large part of the inter-cluster correlations can be treated perturbatively. Third, even though the second-order critical point does not change significantly with the cPT2 correction, the first-order one shifts significantly to $J_2/J_1 \sim 0.62$, which agrees with the $2 \times \infty$ and $4 \times \infty$ extrapolations in cMF. Lastly, we also tried to extrapolate the cPT2 correlation energy to the thermodynamic limit ($L \rightarrow \infty$ and $l \rightarrow \infty$) for some values of $J_2/J_1$. We have to remind the reader that it should approach 0, because clusters of infinite size capture all the energy at the cMF level. Similarly to the cMF analysis, the correlation energy was plotted with respect to the inverse of the cluster size, and at $J_2/J_1=0$ we found a correlation energy of $-0.00185$, which is reasonably close to the expected 0 and again tests the quality of our extrapolation, which is good to about 3 decimal places, similarly to table \ref{tab.1} where the extrapolated results agree to 3 decimal places with the reference results.
\par Regarding cCCSD, we can see that the energy improves compared to the cPT2 energy in the Néel antiferromagnetic phase, but gives higher energies in the other two phases. We think that this is due to the nature of the paramagnetic phase and the nature of the tilted clusters in the collinear antiferromagnetic phase, which suggests that there exist correlations that cannot be captured by using just $2 \times 2$ tiles and a low order coupled-cluster theory. This may imply that triples or quadruples are needed (discussed in the following section). We have to emphasize, however, that at some point cCC should become exact.
\par
\begin{figure}
\centering
\includegraphics[scale=0.7]{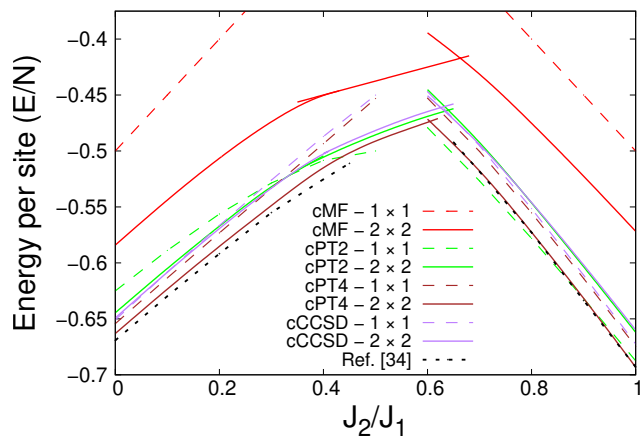}
\caption{A comparison between the $2 \times 2$ and $1 \times 1$ clusters in the thermodynamic limit $L \rightarrow \infty$. The $1 \times 1$ results correspond to the conventional single-reference results. More specifically, cMF-$1 \times 1$ is equal to Hatree-Fock, cPTn-$1 \times 1$ is MPn and cCCSD-$1 \times 1$ is CCSD.}
\label{1x1v2x2}
\end{figure}

 Lastly, regarding cPT4, we decided to compare the $2 \times 2$ with the $1 \times 1$ clusters for all the different methods, mainly for two reasons: 1) This comparison will shed some light on the difference between cPT2 and cCCSD mentioned in the previous paragraph, and 2) The cost of extrapolating cPT4 to the thermodynamic limit $l \rightarrow \infty$ is prohibitive. Our results are summarized in Fig.~\ref{1x1v2x2}. It is easily shown that the $2 \times 2$ results are much better for all the values of $J_2/J_1$ as well as all the different correlation methods. In addition, it is also important to mention that around $J_2/J_1 \approx 0.5$, the $1 \times 1$ results are very poor due to the nature of the mean-field wavefunction (not shown), which is also portrayed in Ref.~\citenum{bishop_phase_1998}, where the authors perform local excitations to very high orders on single-reference wavefunctions. This establishes the fact that clustering, even with small clusters, is meaningful. Moreover, cPT4, which is still considered a low-order perturbation theory, can provide very accurate estimates for the energy even with small clusters, which in fact verifies that small clusters with low-level correlated theories are very useful. Lastly, another observation is that cCCSD being less accurate than cPT2 and cPT4 is not unique to the $2 \times 2$ case, because the results are similar for the $1 \times 1$ case, too. For a more extensive discussion about a comparison of correlated methods we encourage the reader to read the Appendix.

\section{Discussion}\label{5.0}
In section \ref{2.0}, we have described the cluster mean-field
approach to treat strongly-correlated spin systems.
A cMF state is used as a variational ansatz for the ground
state wavefunction, which is guaranteed by construction
to provide better variational estimates than HF when the
size of the cluster is larger than 1. Because of the simple cluster language, a RS-PT scheme can be easily adopted to account for the missing inter-cluster correlations. The results presented in Secs. IV B and IV C provide evidence that a cluster-based approach can (semi)-quantitatively capture the physics of the ground state of the specific benchmark model, the 2D square $J_1 - J_2$ Heisenberg model. Due to the nature of the Heisenberg Hamiltonian, contributions to the second-order energy arise only from two-cluster spin interactions. A significant improvement to the ground state energy is obtained with cPT2, cPT4, as well as cCCSD. We also notice that enlarging the size of the cluster in mean-field calculations is worse than performing a cPT2, cPT4 or a cCCSD calculation; for example, Fig. \ref{nn} shows that cMF with $6 \times 6$ tiles is farther from the exact result than is cPT2 with $2 \times 2$ tiles. The good quality of cPT and cCCSD results suggest that the zero-th order Hamiltonian might be suitable for describing spin lattices in general.
\par In the rest of this section we discuss
possible strategies that can improve the results presented
in this article. The simplest strategy, also discussed in Ref.~\citenum{jimenez-hoyos_cluster-based_2015}, is to use the full Hilbert space (not restricted to a given $S_z$ sector). This will give more variational freedom to the cMF ansatz. This approach, however, requires a Hilbert space of much larger dimension. As regards the cPT2 and the cCCSD methods, the most straightforward way to go beyond those is to use cPTn or cCCSDTQ, etc. To do so, we must truncate the number of states used, because the computational cost will be prohibitive. This truncation scheme could be either based on the local character of the clusters or can be found stochastically. Another possible route could be to exploit locality. For example, we can treat the interaction between nearest-neighbor clusters with cCC and with further clusters with cPT. One more advantage of the cluster-based approaches, is that even though we have used those approaches to study strongly interacting systems, they may be used in other contexts. More specifically, systems which can be effectively represented in terms of weakly interacting fragments of strongly-correlated subsystems can be very efficiently described by cPT2 or cCCSD. Lastly, another route for correlating cMF would be to write the ansatz as a linear combination of different cMFs of different tilings. This has been tried for dimers by Garcia-Bach\cite{garcia-bach_long-range_2000} and has yielded very promising results.

\section{Conclusions}\label{6.0}
In this work, we have used a variational cluster mean-field approach, correlated it with perturbation and coupled-cluster theory, and applied all of them to strongly-correlated spin systems. The optimization of the cluster mean-field state has been carried out with the restriction that the cluster state has well-defined $S_z$ quantum number. The restrictions are imposed in order to preserve total $S_z$ in the full system and facilitate the computation of matrix elements. The cluster product state constitutes an eigenstate of a mean-field (zero-th order) Hamiltonian, which allows us to go beyond mean-field in a perturbative and a coupled-cluster framework. We have presented mean-field, second-order perturbative and coupled-cluster results of the ground state energies of the square 2D Heisenberg model in the thermodynamic limit ($L \rightarrow \infty$). Also, we have presented a relatively accurate extrapolation scheme for thermodynamic limit ($L \rightarrow \infty$ and $l \rightarrow \infty$) energies. In general, we observe that cPT2 and cPT4 energies with small clusters are often better than cMF results with significantly larger ones, and the same applies to cCCSD. Overall, the results of this work suggest that a cluster mean-field approach can provide a good starting point and a path to an efficient description of strongly-correlated systems, while cPT2, cPT4 and cCCSD provide an adequately accurate quantitative description. Several strategies to
improve the mean-field description as well as correlated
approaches built on top of it have been suggested.

\section*{Author Information}
\subsection*{Corresponding Author}
\textbf{Athanasios Papastathopoulos-Katsaros} − Department of Chemistry, Rice University, Houston, Texas 77005, USA; Email: \\ athanasios.papastathopoulos-katsaros@rice.edu

\begin{acknowledgement}
This work was supported by the U.S. Department of Energy,
Office of Basic Energy Sciences, Computational and Theoretical Chemistry Program under Award No. DE-FG02-09ER16053. CAJH is grateful for support from a start-up package at Wesleyan University. G.E.S. acknowledges support as a Welch Foundation Chair (Grant No. C-0036).
\end{acknowledgement}

\appendix
\section{Comparing the correlated methods}\label{app}
\begin{figure}
\centering
\includegraphics[scale=0.7]{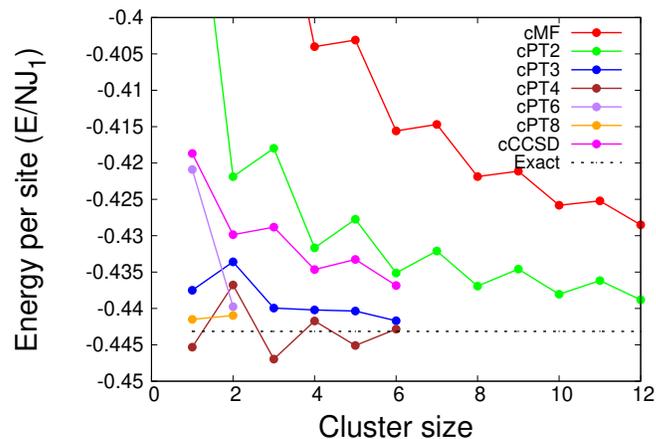}
\caption{Energies per site using different tilings and level of treatment, as well as the extrapolated energy computed as described in section \ref{4.2}.}
\label{pt_order_2}
\end{figure}

One of the main questions in the present paper is if clustering is useful: is it better to remain with small clusters and go to high order in CC or PT, or is it better to use larger clusters? To answer this question, we consider a set of calculations performed on the $J_1 - J_2$ Heisenberg chain. Our results for different cluster sizes and level of correlation at $J_2/J_1 = 0$ are summarized in Fig.~\ref{pt_order_2}. We notice the following important observations. First, with cMF and cPT2 we roughly get a $1/x$ behavior with respect to cluster size, with an even-odd alternation. cPT3 and cPT4 might or might not have that behavior, but we do not have enough points to ensure that. Second, as the cluster size increases, cCCSD approaches cPT2. This is something that we suspect will remain true: cCCSD
(incorrectly) predicts very weak next-nearest neighbor correlations,
and the difference in nearest-neighbor correlations between cPT2 and
cCCSD is small. Lastly, we suggest that with a reasonable cluster size of $1 \times 4$ and a reasonable level of correlation, either cCCSD or cPT3/4, we can get a good balance of accuracy and cost. For example, cPT4 with a cluster size of $1 \times 6$ (error $0.3 \times 10^{-4} J$) is significantly better than LSUB12 from Bishop's work (see Ref.~\citenum{bishop_main}, error $0.8 \times 10^{-4} J$).

\par
\begin{figure}
\centering
\includegraphics[scale=0.7]{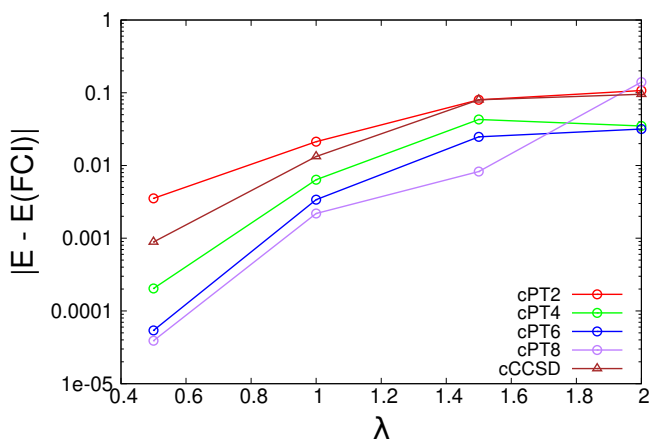}
\caption{Energy error of cCCSD and cPTn with respect to the inter-cluster tuning parameter $\lambda$. Small values of $\lambda$ correspond to better approximations of the exact solution.}
\label{lambda}
\end{figure}

 To shed some light on the differences between cPT2 and cCCSD, we decided to consider two sets of calculations. The first set of calculations considers a toy model on the $J_1-J_2$ Heisenberg chain, similarly to Li's (see Ref.~\citenum{li_block-correlated_2004}) work on block-correlated coupled-cluster for ladder systems. We consider a system where the inter-cluster correlations are scaled by a factor $\lambda$. The exact Hamiltonian is recovered in the limit of $\lambda = 1$ and cMF becomes exact in the limit of $\lambda = 0$. In this work, our proposed methods seek to work very well in the $0 \le \lambda \le 1$ case. By doing that, we can see if the correlated methods improve or deteriorate respectively. More specifically, we use $2 \times 1$ clusters and periodic boundary conditions. The total arrangement corresponds to five $2 \times 1$ clusters, which are equal to a 10x1 system. Our results for $J_2/J_1=0.0$ are summarized in Fig.~\ref{lambda}. The first observation is that for large values of $\lambda$, cPTn and cCCSD break down, which suggests that cPTn and cCCSD are accurate because of relatively small inter-cluster interactions (for small values of $\lambda$ cPTn and cCCSD approach the exact answer very rapidly). In addition, we notice that as expected from single-reference perturbation theory and coupled-cluster, cCCSD significantly improves upon cPT2 and is worse than cPT4, cPT6 and cPT8.
\par
\begin{figure}
\centering
\includegraphics[scale=0.7]{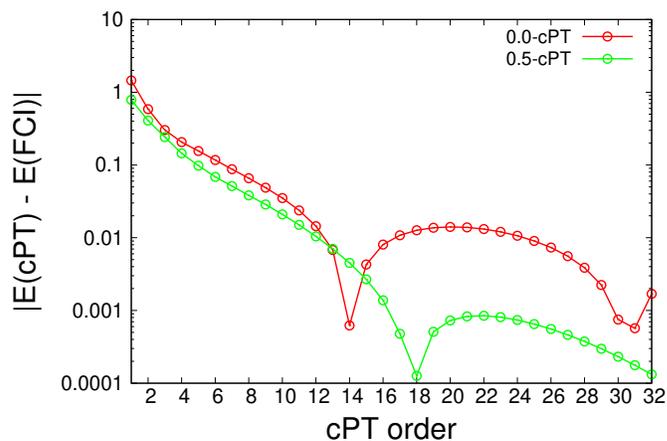}
\caption{A comparison between $J_2/J_1=0.0$ and $J_2/J_1=0.5$ with respect to the cPT order. 0.0-cPT corresponds to cPTn at $J_2/J_1=0.0$. Larger orders of cPT should correspond to better approximations of the exact solution, as long as inter-cluster interactions remain small.}
\label{pt_order}
\end{figure}

 The same observations can be seen with the second set of calculations. This time, we used the same configurations as for the previous one, but only with $\lambda=1.0$ and both $J_2/J_1=0.0$ and $J_2/J_1=0.5$; the difference this time was that the parameter that was changing was the number of orders in the cPT (RS-PT) series. In the ideal case, we want the cPT series to converge to the exact answer with just a few terms and not need many terms to include or diverge completely. In Fig.~\ref{pt_order} we can see that this is truly the case, which suggests that indeed the inter-cluster interactions are relatively small. Lastly, it is important to note that for larger clusters, the weight of inter-cluster compared to intra-cluster correlations is decreased and therefore calculations which use larger clusters are inherently more weakly correlated. This explains the better performance of cPT2 (and possibly cCCSD) in those scenarios.

%\section*{Data availability}
%The data that support the findings of this study are available from the %corresponding author upon reasonable request.

%\nocite{*}

%-------------------------------------
\bibliography{cc}
%-------------------------------------

%-------------------------------------
%-------------------------------------

\onecolumn

\section*{Supporting Information}
Table showing the energy per site obtained by cMF, cPT2 and cCCSD calculations at different J2/J1 for a 5 × 5 system in terms of clusters with periodic boundary conditions.

\begin{table}[]
\begin{tabular}{|l|llllll|}
\hline
$J_2/J_1$          & \multicolumn{1}{l|}{cMF - 2 x 2} & \multicolumn{1}{l|}{cMF - 4 x 4} & \multicolumn{1}{l|}{cPT2 - 2 x 2} & \multicolumn{1}{l|}{cPT2 - 4 x 4} & \multicolumn{1}{l|}{cPT4 - 2 x 2} & cCCSD - 2 x 2 \\ \hline
Néel order         & \multicolumn{6}{r|}{}                                                                                                                                                                           \\ \hline
0.0                & \multicolumn{1}{l|}{-0.584053}   & \multicolumn{1}{l|}{-0.624400}   & \multicolumn{1}{l|}{-0.644554}    & \multicolumn{1}{l|}{-0.656360}    & \multicolumn{1}{l|}{-0.663120}    & -0.649224     \\ \hline
0.1                & \multicolumn{1}{l|}{-0.543854}   & \multicolumn{1}{l|}{-0.585287}   & \multicolumn{1}{l|}{-0.604984}    & \multicolumn{1}{l|}{-0.616246}    & \multicolumn{1}{l|}{-0.623431}    & -0.608385     \\ \hline
0.2                & \multicolumn{1}{l|}{-0.506536}   & \multicolumn{1}{l|}{-0.548614}   & \multicolumn{1}{l|}{-0.567144}    & \multicolumn{1}{l|}{-0.577798}    & \multicolumn{1}{l|}{-0.585243}    & -0.569349     \\ \hline
0.3                & \multicolumn{1}{l|}{-0.473925}   & \multicolumn{1}{l|}{-0.515506}   & \multicolumn{1}{l|}{-0.532011}    & \multicolumn{1}{l|}{-0.541791}    & \multicolumn{1}{l|}{-0.548778}    & -0.533178     \\ \hline
0.4                & \multicolumn{1}{l|}{-0.450314}   & \multicolumn{1}{l|}{-0.488336}   & \multicolumn{1}{l|}{-0.504373}    & \multicolumn{1}{l|}{-0.510434}    & \multicolumn{1}{l|}{-0.514966}    & -0.502659     \\ \hline
Paramagnetic order & \multicolumn{6}{l|}{}                                                                                                                                                                           \\ \hline
0.4                & \multicolumn{1}{l|}{-0.450000}   & \multicolumn{1}{l|}{-0.487971}   & \multicolumn{1}{l|}{-0.505509}    & \multicolumn{1}{l|}{-0.510478}    & \multicolumn{1}{l|}{-0.514966}    & -0.502273     \\ \hline
0.5                & \multicolumn{1}{l|}{-0.437500}   & \multicolumn{1}{l|}{-0.469097}   & \multicolumn{1}{l|}{-0.485590}    & \multicolumn{1}{l|}{-0.487792}    & \multicolumn{1}{l|}{-0.491195}    & -0.481673     \\ \hline
0.6                & \multicolumn{1}{l|}{-0.425000}   & \multicolumn{1}{l|}{-0.452088}   & \multicolumn{1}{l|}{-0.469097}    & \multicolumn{1}{l|}{-0.470728}    & \multicolumn{1}{l|}{-0.474357}    & -0.464788     \\ \hline
Collinear order    & \multicolumn{6}{l|}{}                                                                                                                                                                           \\ \hline
0.6                & \multicolumn{1}{l|}{-0.394376}   & \multicolumn{1}{l|}{-0.430190}   & \multicolumn{1}{l|}{-0.445175}    & \multicolumn{1}{l|}{-0.457787}    & \multicolumn{1}{l|}{-0.471560}    & -0.450628     \\ \hline
0.7                & \multicolumn{1}{l|}{-0.431394}   & \multicolumn{1}{l|}{-0.470243}   & \multicolumn{1}{l|}{-0.492452}    & \multicolumn{1}{l|}{-0.502625}    & \multicolumn{1}{l|}{-0.517835}    & -0.494262     \\ \hline
0.8                & \multicolumn{1}{l|}{-0.475245}   & \multicolumn{1}{l|}{-0.518060}   & \multicolumn{1}{l|}{-0.546428}    & \multicolumn{1}{l|}{-0.556238}    & \multicolumn{1}{l|}{-0.572628}    & -0.546040     \\ \hline
0.9                & \multicolumn{1}{l|}{-0.522516}   & \multicolumn{1}{l|}{-0.569710}   & \multicolumn{1}{l|}{-0.603393}    & \multicolumn{1}{l|}{-0.613883}    & \multicolumn{1}{l|}{-0.631654}    & -0.601842     \\ \hline
1.0                & \multicolumn{1}{l|}{-0.571747}   & \multicolumn{1}{l|}{-0.623527}   & \multicolumn{1}{l|}{-0.662009}    & \multicolumn{1}{l|}{-0.673711}    & \multicolumn{1}{l|}{-0.693110}    & -0.659863     \\ \hline
\end{tabular}
\caption{Energy per site obtained by cMF, cPT2 and cCCSD calculations at different $J_2/J_1$ for a $5 \times 5$ system in terms of clusters with periodic boundary conditions.}
\label{si.1}
\end{table}

\end{document}